\documentclass[aps,twocolumn,showpacs]{revtex4-1}
\usepackage{graphicx}
\usepackage{amsfonts}
\usepackage{amssymb}
\usepackage{amsbsy}
\usepackage{amsmath}
\usepackage{mathrsfs}
\usepackage{latexsym}
\usepackage{natbib}
\usepackage{bm}
\usepackage{subfigure}
\usepackage{color}
\usepackage{dsfont}
\usepackage{dirtytalk}
\usepackage{hyperref}

\newcommand{\Hq}{\mathcal{H}}

\newcommand{\hs}{\mathcal{H}_{\text{\tiny S}}}
\newcommand{\hl}{\mathcal{H}_{\text{\tiny L}}}
\newcommand{\hsl}{\mathcal{H}_{\text{\tiny SL}}}

\newcommand{\nn}{\nonumber}

\newcommand{\so}{\mathcal{S}}

\def\xx{\mathbf{x}}

\def\x{\mathbf{x}}

\def\rs{r_s}

\begin{document}

%\title{Resonance Casimir-Polder interaction in Schwarzschild spacetime for a massive scalar field}

\title{How the mass of a scalar field influences Resonance Casimir-Polder interaction in Schwarzschild spacetime }

\author{Arpan Chatterjee}
\email{ac17rs016@iiserkol.ac.in}

\author{Saptarshi Saha}
\email{ss17rs021@iiserkol.ac.in}

\author{Chiranjeeb Singha}
\email{cs12ip026@iiserkol.ac.in}

\affiliation{ Department of Physical Sciences, 
Indian Institute of Science Education and Research Kolkata,
Mohanpur - 741 246, WB, India }
 
\pacs{04.62.+v, 04.60.Pp}

\date{\today}

\begin{abstract}
 
\emph{Resonance Casimir-Polder interaction(RCPI)} occurs in nature when one or more atoms are in their excited states and the exchange of real photon is involved between them due to vacuum fluctuations of the quantum fields. 
 In recent times, many attempts have been made to show that the curved spacetime such as the \emph{de-Sitter spacetime} can be separated from a \emph{thermal Minkowski spacetime} using \emph{RCPI}. Motivated from these ideas, here we study the \emph{RCPI} between two atoms that interact with a massive scalar field in Schwarzschild spacetime provided the atoms are placed in the near-horizon region. Subsequently, we use the tool of the \emph{open quantum system} and calculate the Lamb shift of the atomic energy level of the entangled states. We show that the behavior of \emph{RCPI} modifies depending on the mass of the scalar field. In the high mass limit, the interaction becomes short-range and eventually disappears beyond a characteristic length scale of $1/m$, where $m$ is the mass of the scalar field.

\end{abstract}
\maketitle

\section{Introduction} \label{introduction}

The phenomena of \emph{Casimir Polder interaction (CPI)} occurs between an atom and a conducting plate due to vacuum fluctuation of the quantized fields \cite{casimir_influence_1948,fiftyyears,dalvit_casimir_2011}. It has major applications from the field of quantum electrodynamics (QED) to the signature of gravity \cite{jaffe_casimir_2005,berman_cavity_1994,tian_detecting_2016}. \emph{CPI} mimics van der Waals force in non-relativistic regime where a dispersive force is acting in few nanometer length scale \cite{macdowell_surface_2019,klimchitskaya_casimir_2015}. In physical chemistry it has also huge implications in atomic adsorption and desorption in the thermally excited surfaces \cite{lin_impact_2004}. Experimental realization exists in the case of the motion of Bose-Einstein condensate (BEC) under surface \emph{CPI} \cite{harber_measurement_2005}. The other  successful efforts  have 
been made to probe more complicated contexts like detecting   
spacetime curvature 
\cite{Tian:2014hwa,tian_detecting_2016,PhysRevA.88.064501}, Unruh effect 
\cite{PhysRevA.76.062114,Marino:2014rfa,Rizzuto:2016ijj}, Hawking radiation of a black hole and checking 
thermal and nonthermal scaling in a black hole spacetime \cite{Menezes:2017akp,Yu:2008zza}. One can show that the background spacetime and the relativistic motion of 
interacting systems can modify the \emph{CPI} 
\cite{PhysRevA.76.062114,Marino:2014rfa,Rizzuto:2016ijj,Tian:2014hwa, 
tian_detecting_2016,PhysRevA.88.064501,Yu:2008zza,Menezes:2017akp,Zhou:2017axh,PhysRevD.97.105030,PhysRevD.79.044027,PhysRevD.88.104003,Salton_2015}.

\emph{CPI} generates correlations between multiple quantum systems \cite{felicetti_dynamical_2014,busch_quantum_2014,Cirone_2007}. Two uncorrelated atom also can be entangled if they both share common environment \cite{braun,benatti_environment_2003}. 
Recently Lindblad master equation approach is \cite{PhysRevA.70.012112} successfully used to explain the interatomic correlations at Schwarzschild spacetime where vacuum fluctuation of the quantum field takes the vital role to create the entanglement between them. This interatomic correlations are the source of \emph{RCPI} \cite{tian_detecting_2016,singha_remarks_2019}. Subsequently it has been shown that an uniformly accelerated atom behaves like a system in a thermal bath (\emph{Thermalization theorem}) \cite{Takagi_1986}. Therefore, \emph{RCPI} is exactly equal to the second order shift (Lamb shift) \cite{breuer} due to system-field interaction Hamiltonian   \cite{laliotis_casimir-polder_2015,hollberg_measurement_1984,tian_detecting_2016}.

In this paper following the recent works, \cite{tian_detecting_2016,singha_remarks_2019},  we theoretically investigate the \emph{RCPI} between two atoms that interact with a massive scalar field in Schwarzschild spacetime. Two atoms are initially uncorrelated and interact individually with the scalar field. The cross-terms of the two individual system-field coupling Hamiltonian in the second-order of the quantum master equation will give rise to the entanglement between the atoms \cite{benatti_environment_2003}. Here we show how the interatomic correlations between the atoms depend on the mass of the scalar field by calculating the Lamb shift due to system-field coupling.

We organize the paper in the following way. In section-\ref{dyn} we give a general description of \emph{open quantum system} approach-GKSL (Gorini-Kossakowski-Sudarshan-Lindblad) formalism  \cite{lindblad_generators_1976,kossakowski_quantum_1972,book:9622} for the two-atom system then discuss the common environment effect and finally compute the shifts of the energy level of the entangled states due to Lamb shift Hamiltonian. In section- \ref{horizon} we review the form of the Schwarzschild metric in the near horizon and find the expression of the two-point correlation function for a massive scalar field. Recently it has been shown that, for Schwarzschild spacetime, beyond a characteristic length scale which is proportional to the inverse of the surface gravity $\kappa$, the \emph{RCPI} between two entangled atoms is characterized by a $1/L^2$  power-law provided the atoms are located close to the horizon \cite{singha_remarks_2019}. 
However, the length scale limit beyond the characteristic value is not compatible with the local flatness of the spacetime. A massless scalar field has been considered in the ref. \cite{singha_remarks_2019} to calculate \emph{RCPI}.  
In section-\ref{lamb} we compute \emph{RCPI} between two atoms that interact with a massive scalar field and point out its dependence on the mass term of the scalar field. Subsequently, we show that the \emph{RCPI} becomes short-range and eventually disappears beyond a characteristic length scale of $1/m$, $m$ being the mass of the scalar field.

\section{Dynamics of a two-atom system} \label{dyn}

We consider a system of two-atom weakly interacts with a common environment but they do not have any mutual interaction. Here we choose the environment to be a quantized massive scalar field. This is a general description of the \emph{open quantum system} \cite{atom-photon}, therefore, we follow the formalism of the quantum master equation \cite{breuer}. The Hamiltonian of the full system can be expressed as,
\begin{eqnarray}
\Hq=\hs^0+\hl^0+\hsl~.
\label{eq:1}
\end{eqnarray}
We assume the atoms have the same Zeeman levels and throughout the paper, we use the natural units, $\hbar=c=1$. The free Hamiltonian of the system can be written as $\hs^0=\hs^{01}+\hs^{02}=\frac{1}{2}\omega_0 \sigma_3^{(1)}+\frac{1}{2}\omega_0 \sigma_3^{(2)}$ (The superscript in the Pauli matrices represent the atom number). Here $\sigma_i$ are the Pauli matrices, $\omega_0$ is the frequency of Zeeman levels and we defined $\vert g \rangle, \vert e \rangle$ as the corresponding ground and excited state of the atoms. The free Hamiltonian of the scalar field $\hl^0$ can be written as \cite{schroeder_introduction_1995},
\begin{eqnarray}
\hl^0=\int \frac{d^3 k}{(2\pi)^3}\omega_k a^{ \dagger}(k) a(k)~, 
\label{eq:2}
\end{eqnarray}
where $\omega_k=\sqrt{k^2+m^2}$, the frequency of the scalar field. Here m is the mass of the scalar field and $a^{ \dagger}, a$ are the creation and annihilation operator of the quantized field respectively. The coupling Hamiltonian between the atom-scalar field can be defined as \cite{PhysRevA.70.012112},
\begin{eqnarray}
\hsl = \lambda \sum\limits_{\mu=0}^{3} \big[ \sigma_{\mu }^{(1)} \otimes \phi(x_1)   +  \sigma_{\mu}^{(2)} \otimes \phi(x_2) \big]~,
\label{eq:3}
\end{eqnarray} 
where $\lambda$ is the coupling constant and $\phi$ represents the scalar field. Initially, the atoms and the scalar field are separated from each other. Therefore, the initial density matrix can be written as $\rho(0)= \rho_s(0) \otimes \vert 0 \rangle \langle 0 \vert$ (``Born-Markov" approximation \cite{atom-photon}). Here $\vert 0 \rangle$ is the vacuum state of the massive scalar field and $\rho_s(0)$ is the initial density matrix of the system. For a closed system, ``Von-Neumann-Liouville" equation is used to describe the full dynamics, 
\begin{eqnarray} 
\label{neuman}
\frac{d \rho(t)}{dt}&=& -i [\hs^0+\hl^0+\hsl, \rho (t)].
%\label{propa}
\end{eqnarray}
Here, $\rho(t)$ is the total density matrices.
Starting from the equation (\ref{neuman}) we perturbatively expand the time evolution operator \big($U(t,t_0)=\mathcal{T}\exp \{-i \int^{t}_{t_0} dt^{\prime} (\hs^0+\hl^0+\hsl)\}$\big) to the second order of coupling Hamiltonian \cite{breuer} and taking trace over field degrees of freedom, we get the quantum master equation, which is also known as GKSL equation \cite{lindblad_generators_1976,kossakowski_quantum_1972,book:9622}. The quantum master equation in the lab-frame with a proper-time $\tau$ can be expressed as \cite{breuer},
\begin{eqnarray}
\frac{d \rho_s(\tau)}{d\tau}= -i[\hs^0 + \mathcal{H}_{lamb},\rho_s(\tau)]+\mathcal{L}\big(\rho_s(\tau)\big)~. 
\label{eq:4}
\end{eqnarray}
Here, $\mathcal{H}_{lamb}$ is the Lamb-shift Hamiltonian that leads to the renormalization of Zeeman Hamiltonian and $\mathcal{L}\big(\rho_s(\tau)\big)$ is the dissipator of the master-equation which can be written as \cite{PhysRevA.70.012112},
\begin{eqnarray}
\mathcal{L}\big(\rho_s\big) &=& \sum \limits_{a,b=1}^{2} \sum \limits_{j,k=1}^3 \gamma^{ab}_{jk}\Big(\sigma _b ^k\rho_s \sigma _a ^j  
%\qquad\qquad\,\,\,\qquad
-\frac{1}{2}\{ \sigma _a ^j \sigma _b ^k ,\rho_s \} \Big)~,\\
\label{eq:5}
\mathcal{H}_{lamb}&=& -\frac{i}{2} \sum \limits_{a,b=1}^{2} \sum \limits_{j,k=1}^3 \mathcal{S}_{jk}^{ab}  \sigma _a ^j \sigma _b ^k~,
\label{eq:6}
\end{eqnarray}
where $\mathcal{S}_{jk}^{ab}$ and $ \gamma^{ab}_{jk}$ depend on the Fourier transforms of the two-point correlation functions of the scalar field. The shift terms and dissipative terms are coming from second-order terms of atom-field coupling Hamiltonian and they are \emph{Kramer-kronig} pair to each-other \cite{atom-photon}.
 As we are interested in the shift between the energy levels of the two-atom system, our main focus is to demonstrate the Lamb-shift term and not to think about decay terms. However, the decay terms are responsible for taking the two-atom system to the equilibrium configuration, which is the \emph{Bekenstein-Hawking temperature} \cite{Takagi_1986} of the massive scalar field.
 Lamb shift can be calculated from the Hilbert transforms of the Fourier transforms of the two-point functions which are shown below,
\begin{eqnarray}
\mathcal{K}^{ab}(\omega_0) = \frac{\mathcal{P}}{\pi i}\int \limits ^{\infty}_{-\infty}d\omega \frac{\mathcal{G}^{ab}(\omega)}{\omega-\omega_0}~.
\label{eq:7}
\end{eqnarray} 
Here $\mathcal{P}$ denotes the principal value. $\mathcal{G}^{ab}(\omega)$ are the Fourier transforms of the two-point correlation functions of the scalar field. The Fourier transforms of 
the two-point functions are given by, 
\begin{eqnarray}
 \mathcal{{G}}^{ab}(\omega)&=&\int\limits^{\infty}_{-\infty} d \Delta \tau 
~e^{i\omega \Delta \tau}~{G}^{ab}(\Delta\tau)~,\\
\label{eq:8}
\text{where,} \,\,\,\, G^{a b}(\Delta \tau)&=&\langle{\Phi}(\tau,\xx_{a}){\Phi}(\tau',\xx_{b})\rangle~,
\label{eq:9}
\end{eqnarray}
and $\Delta \tau = (\tau-\tau')$. The exact form of $\mathcal{S}^{ ab } _ { jk }$ can be written as \cite{tian_detecting_2016}, 
\begin{eqnarray}
 \mathcal{S}^{ ab } _ { jk }=A^{a b}\delta_{jk}-i B^{a 
b}\epsilon_{jkl}\delta_{3 l}-A^{a b}\delta_{3j}\delta_{3k}~,
\label{eq:10}
\end{eqnarray}
where the terms $A^{a b}$ and $B^{a b}$ are given by,
\begin{eqnarray}
 A^{a b}&=&\frac{\lambda^2}{4}\left[\mathcal{K}^{ab} 
(\omega_0)+\mathcal{K}^{ab}(-\omega_0)\right],
\label{eq:11}\\
B^{ab}&=&\frac{\lambda^2}{4}\left[\mathcal{K}^{ab} 
(\omega_0)-\mathcal{K}^{ab}(-\omega_0)\right]
\label{eq:12}.
\end{eqnarray}

\subsection{Entanglement between two atoms through a massive scalar field}\label{entang}
%\subsection{Common environment effect between the two atoms}
Two atoms are initially uncorrelated but due to interaction with the massive scalar field, the energy levels become correlated. So, there is a formation of field-induced entanglement between the atoms which depends on the two-point correlation functions \cite{benatti_environment_2003}. The two-point functions can be computed along the trajectories of the atoms thus they depend on the spacetime background.  Now we aim to calculate the expectation value of $\mathcal{H}_{lamb} $ in the entangled states to investigate the effect of interatomic correlations between two atoms in Schwarzschild spacetime. Here we take the symmetric and anti-symmetric state of the dipolar-coupled Hamiltonian for simplicity because other states are unchanged due to interatomic correlations \cite{FICEK2002369}. The symmetric and anti-symmetric states of two atom-system are given by, 
\begin{eqnarray}
\vert E \rangle = \frac{\vert e_1 \rangle\vert g_2 \rangle + \vert g_1 \rangle\vert e_2 \rangle}{\sqrt{2}}~;\quad \vert A \rangle = \frac{\vert e_1 \rangle\vert g_2 \rangle - \vert g_1 \rangle\vert e_2 \rangle}{\sqrt{2}}~.\nn
\end{eqnarray} 
One can find the expressions for energy level shifts as \cite{tian_detecting_2016},
\begin{eqnarray}
&&\delta E_{S_{LS}}=\langle
E|\mathcal{H}_{lamb}|E\rangle\nn\\
&&=-\frac{i}{2}\left[\sum^{3}_{j=1}\left(\so^{12}_{
jj}+\so^{21}_{jj}+\so^{11}_{jj}+\so^{22}_{jj}\right)-2(\so^{12}_{33}+\so^{ 21 } _ { 33 } 
) \right],
\nonumber\\
&&\delta E_{A_{LS}}=\langle
A|\mathcal{H}_{lamb}|A\rangle\nn\\
&&=\frac{i}{2}\left[\sum^{3}_{j=1}\left(\so^{12}_{
jj}+\so^{21}_{jj}-\so^{11}_{jj}-\so^{22}_{jj}\right)\right].
\label{shift}
\end{eqnarray}

Here $\delta E_{S_{LS}}$ stands for the second-order energy level shift of the 
symmetric state and $\delta E_{A_{LS}}$ stands for corresponding energy level 
shift of the antisymmetric state. In the next section we
compute these shifts of the energy level of the entangled states for a massive scalar field in Schwarzschild spacetime when the atoms are located close to the horizon.

\section{Schwarzschild metric in near horizon region}\label{horizon}

Here we consider (3+1) dimensional Schwarzschild spacetime to compute the 
\emph{RCPI} between two entangled atoms interact with a massive scalar field when the atoms are located close to the 
horizon. The Schwarzschild spacetime is described by the line element,
\begin{equation}\label{SchwarzschildMetric01}
d{s}^2 = - f(r) dt^2 + f(r)^{-1} dr^2 
+ r^2 d\theta^2 + r^2 \sin^2\theta d\phi^2 ~,
\end{equation}
where, $f(r) = \left(1- r_s /r\right)$ and $r_s = 2 G M$ is the Schwarzschild 
radius related to the line element. A proper distance $l$ from the horizon 
to a radial distance $r$ is defined by the formula \cite{Ydri:2017oja},
\begin{eqnarray}
&&l=\sqrt{r(r-r_s)}+r_s \sinh^{-1}(\sqrt{\frac{r}{r_s}-1})~.
\end{eqnarray}
In terms of $l$, the Schwarzschild metric (\ref{SchwarzschildMetric01}) becomes
\begin{equation}\label{SchwarzschildMetric02}
d{s}^2 = - f(r) dt^2 + d l^2
+ r^2(l) d\theta^2 + r^2(l) \sin^2\theta d\phi^2 ~,
\end{equation}
where, $f(r)= \left(1- r_s /r(l)\right)$. Now near the horizon, where 
$r=r_s+\delta$, $l=2 \sqrt{r_s\delta}$ and on a small angular region which is around 
$\theta \sim 0$, the new coordinates have been defined as \cite{singha_remarks_2019},  
\begin{eqnarray}\label{Kruskal}
X_1&=&l\cosh\frac{t}{2 r_s}~;~~ T=l\sinh\frac{t}{2 r_s}~;\nn\\
X_2&=& r_s\theta \cos\phi~;~~X_3=r_s \theta \sin\phi~.
\end{eqnarray}
Using this coordinates (\ref{Kruskal}), the line element (\ref{SchwarzschildMetric01}) becomes \cite{singha_remarks_2019},
\begin{equation}\label{Minkowski1}
 d{s}^2=-d T^2+d X_1^2+d X_2^2+d X_3^2~,
\end{equation}
which expresses the Minkowski spacetime \cite{Ydri:2017oja, Polchinski:2016hrw,Dabholkar:2012zz, 
Goto:2018ijz,singha_remarks_2019}.

\subsection{Two point correlation function for the massive scalar field}
In the position space, using the coordinates of the inertial metric 
(\ref{Minkowski1}), the two-point function for a massive scalar field of mass $m$ can be expressed as \cite{schroeder_introduction_1995},
\begin{eqnarray}
G(x,x^{\prime})&\equiv&\langle 0 \vert \hat{\Phi} (x) \hat{\Phi}(x^{\prime})\vert 0\rangle =\langle 0|\hat{\Phi}(T,\x) \hat{\Phi}(T',\x')|0\rangle\nn\\&=& \int \frac{d^4 k}{(2\pi)^{\frac{3}{2}}} \delta(k^2-m^2)e^{-ik(x-x^{\prime})}~.
\end{eqnarray}
After integrating the above expression, the two-point function for the scalar field can be written as \cite{birrell_quantum_1982},
\begin{eqnarray}\label{cor}
G(x-x^{\prime})=-\frac{im}{4\pi^2}\frac{K_1(im \sqrt{(T-T^{\prime}-i \epsilon)^2-(\x-\x^{\prime})^2})}{\sqrt{(T-T^{\prime})^2-(\x-\x^{\prime})^2}}~.\nn\\
\end{eqnarray}
Here $K_1$ is the modified Bessel function of the second kind, $(\x-\x^{\prime})^2= 
(X_1-X_1')^2+(X_2-X_2')^2+(X_3-X_3')^2$  and $\epsilon$ is a small, positive parameter which is introduced to 
evaluate two-point function. In the limit $mR<<1$ the function behaves like the two-point correlation function (\ref{cor}) for a massless scalar field case \cite{chowdhury_unruh-dewitt_2019}. In this limit, the expression of two-point function is shown below,
\begin{eqnarray}
G(x-x^{\prime})&=&-\frac{1}{4 \pi^2 R^2}~.
\end{eqnarray} 
On the other hand in the high mass limit, $mR>>1$, the correlation function (\ref{cor}) is given by \cite{chowdhury_unruh-dewitt_2019},
\begin{eqnarray}
G(x-x^{\prime})&=& \frac{1}{4 \pi R}\Big(\frac{i m}{2 \pi R}\Big)^\frac{1}{2}e^{-im R}~,\nn\\
\text{where,} \qquad R &=& \sqrt{(T-T^{\prime})^2-(\x-\x^{\prime})^2}~.
\end{eqnarray}
Using equation (\ref{Kruskal}) $R$ can be written as \cite{singha_remarks_2019},
\begin{eqnarray}
R= 2l \sinh \frac{ (\tau-\tau^{\prime})}{2 l}~.
\end{eqnarray}
Now to avoid divergence at $(\tau-\tau^{\prime}=0), R$ is transformed by the following transformation ($\tau$ is the proper time),
\begin{eqnarray}
R= 2l \sinh \frac{ (\tau-\tau^{\prime})}{2 l} + i \epsilon~.
\end{eqnarray}
Here $1/l$ behaves as the acceleration of the system and $\epsilon$ is a small positive quantity. So,
\begin{eqnarray}
\vert G(x-x^{\prime}) \vert &=& \Big(\frac{m}{32 \pi^3 \vert R \vert ^3}\Big)^{\frac{1}{2}} e^{-m \epsilon}~,\nn\\
 \text{where} \qquad \vert R \vert ^2 &=& \epsilon^2 + \Big( 2l \sinh \frac{ (\tau-\tau^{\prime})}{2 l}\Big)^2.
\end{eqnarray}
Using Lebesgue's bounded convergence theorem the Fourier transform of correlation function goes to zero for the large mass limit \cite{chowdhury_unruh-dewitt_2019,Takagi_1986}, which also indicates in high mass limit \emph{RCPI} may disappear.

\section{\emph{RCPI} for the two-atom system} \label{lamb}
Two static atom placed in two different position $\to (r,\theta,\phi)\,\, \text{and}\,\, (r,\theta^{\prime},\phi) $ which are close to the horizon. Here $\theta$ and $\theta^{\prime}$ are assumed to be small. The response function for a detector (atom) per unit time $(T_0)$  is given by \cite{birrell_quantum_1982},
\begin{eqnarray}
\frac{\mathcal{F}(\omega)}{T_0}&=&\int \limits_{0}^{\infty} \frac{dk\, k^2}{2\pi \sqrt{k^2+m^2}}\vert \beta_k \vert^2 \delta(\omega- \sqrt{k^2+m^2})~.
\end{eqnarray}
 From the analogy of \emph{thermalization  theorem} \cite{Takagi_1986} $\vert\beta_k^2\vert$  is the number of particle with mode $k$ of the isotropic bath with temperature $\mathcal{T_H}=\frac{1}{2 \pi l}$ in the static space time. 
 Following the above discussion, the Fourier transforms of the two-point correlation functions (\ref{eq:8}) for
these two spacetime points can be written as,
\begin{eqnarray}
\mathcal{G}^{11}(\omega)&=&\mathcal{G}^{22}(\omega)\nn\\&=&\frac{1}{2 \pi}\frac{\Omega(\omega,m)}{1-e^{-2 \pi l \Omega(\omega,m)}}~,\nn\\
\mathcal{G}^{12}(\omega)&=&\mathcal{G}^{21}(\omega)\nn\\&=&\frac{1}{2 \pi}\frac{\Omega(\omega,m)}{1-e^{-2 \pi l \Omega(\omega,m)}} g(\Omega(\omega,m),L/2)~.
\label{gg}
\end{eqnarray}
Here we define $\Omega(\omega,m)=\sqrt{\omega^2-m^2}\,\,\theta(\omega-m)$. For $m=0$ the expression exactly matches with the ref. \cite{singha_remarks_2019}. From the mathematical point of view $\omega$ is replaced with $\sqrt{\omega^2-m^2}\,\,\theta(\omega-m)$ for this case. 
We also define $g\big(\Omega(\omega,m),z \big)= \frac{\sin[2l\,\Omega(\omega,m)\sinh^{-1}(z/l)]}{2z \,\Omega(\omega,m) \sqrt{1+z^2/l^2}}$ and $L$ denotes the proper distance between $(r,\theta,\phi)$ and $(r,\theta^{\prime},\phi)$. 
Calculating the Hilbert transforms (\ref{eq:7}) of the above functions (\ref{gg}) the coefficients (\ref{eq:11},\ref{eq:12}) can be written as,
\begin{eqnarray}
A_{1}&=&\frac{\lambda^2 \mathcal{P}}{8 \pi^2 i}\int^{\infty}_{-\infty} d\omega\, \Omega(\omega,m)
\left(\frac{1}{\omega-\omega_0}+\frac{1}{\omega+\omega_0}\right)\times\nn\\
&&\qquad  \qquad \qquad \qquad\qquad\qquad\frac{
1 } { 1-e^{-2 \pi l\Omega(\omega,m)}}~,\nonumber\\
B_{1}&=&\frac{\lambda^2 \mathcal{P}}{8
\pi^2 i}\int^{\infty}_{-\infty} d \omega\, 
\Omega(\omega,m)
\left(\frac{1}{\omega-\omega_0}-\frac{1}{\omega+\omega_0}\right)\times \nn\\
&& \qquad \qquad \qquad\qquad\qquad\frac{
1 } {
1-e^{-2 \pi 
l\Omega(\omega,m)}}~,\nonumber\\
A_{2}&=&\frac{\lambda^2 \mathcal{P}}{8
\pi^2 i}\int^{\infty}_{-\infty} d \omega \,
\Omega(\omega,m)
\left(\frac{1}{\omega-\omega_0}+\frac{1}{\omega+\omega_0}\right) \times\nn\\
&&\qquad \qquad   \frac{
1 } {
1-e^{-2 \pi 
l\Omega(\omega,m)}}g(\Omega(\omega,m),L/2)~,\nonumber\\
B_{2}&=&\frac{\lambda^2 \mathcal{P}}{8
\pi^2 i}\int^{\infty}_{-\infty} d \omega \,
\Omega(\omega,m)
\left(\frac{1}{\omega-\omega_0}-\frac{1}{\omega+\omega_0}\right) \times \nn\\
&&\qquad \qquad   \frac{
1 } {
1-e^{-2 \pi 
l\Omega(\omega,m)}}g(\Omega(\omega,m),L/2)~.
\label{aa}
\end{eqnarray}
Here, $A^{11}=A^{22}=A_1, ~A^{12}=A^{21}=A_2,~ B^{11}=B^{22}=B_1,~ B^{12}=B^{21}=B_2$.
Put the coefficients (\ref{aa}) in the following equations (\ref{eq:10}),
\begin{eqnarray}\label{kaka}
 \mathcal{S}^{ 11 } _ { jk }=\mathcal{S}^{ 22 } _ { jk } =A_{1}\delta_{jk}-i 
B_{1}\epsilon_{jkl}\delta_{3 l}-A_{1}\delta_{3j}\delta_{3k}~,\nn\\
\mathcal{S}^{ 12 } _ { jk }=\mathcal{S}^{ 21 } _ { jk } =A_{2}\delta_{jk}-i 
B_{2}\epsilon_{jkl}\delta_{3 l}-A_{2}\delta_{3j}\delta_{3k}~,
\end{eqnarray}
 the shifts of the energy level of the symmetric state and the anti-symmetric state (\ref{shift}) of the two-atom system  are given by,
\begin{eqnarray}
\delta E_{S_{LS}}&=& -\frac{\lambda^2 }{4
\pi^2 }\int^{\infty}_{0} d \omega \,
\Omega(\omega,m)\left(\frac{1}{\omega-\omega_0}+\frac{1}{\omega+\omega_0}
\right)\times\nn\\
&&\qquad \qquad \qquad \left[ g(\Omega(\omega,m),L/2)+1\right]~,\nn\\
\delta E_{A_{LS}}&=&\frac{\lambda^2 }{4
\pi^2 }\int^{\infty}_{0} d \omega \, 
\Omega(\omega,m)\left(\frac{1}{\omega-\omega_0}+\frac{1}{\omega+\omega_0}
\right) \times \nn\\
&&\qquad \qquad \qquad \left[ g(\Omega(\omega,m),L/2)-1\right]~.
\end{eqnarray}
These shifts depend on the proper length  $L$. To calculate the Casimir-Polder force between the atoms we need to take a derivative with respect to $L$ \citep{tian_detecting_2016}. So, we neglect the terms which do not depend on $L$. From the above discussion, the Lamb shift of the symmetric and anti-symmetric state of the two-atom system is then given by,
\begin{eqnarray}
\delta E_S &=& -\frac{\lambda^2}{4 \pi^2}\int \limits_0^{\infty} d\omega \,\Omega(\omega,m)\Big(\frac{1}{\omega-\omega_0}+\frac{1}{\omega+\omega_0}\Big)\times \nn\\
&& \qquad \qquad \qquad   g(\Omega(\omega,m), L/2)~,\nn\\
\delta E_A &=& \frac{\lambda^2}{4 \pi^2}\int \limits_0^{\infty} d\omega\,\Omega(\omega,m)\Big(\frac{1}{\omega-\omega_0}+\frac{1}{\omega+\omega_0}\Big) \times \nn\\
&& \qquad \qquad \qquad  g(\Omega(\omega,m), L/2)~.
\end{eqnarray}
After substituting $\omega^2-m^2=z^2$, the integral form look like,
\begin{eqnarray}\label{int}
\mathcal{I}=\mp\frac{\lambda^2}{2 \pi^2}\int \limits_0^{\infty} dz \frac{z}{z^2+m^2-\omega_0^2}\,\eta \sin (\alpha z)~,
\end{eqnarray} 
where $\eta=\frac{1}{2L \sqrt{1+\frac{L^2}{4l^2}}}$ and $\alpha=2l\sinh^{-1}(\frac{L}{2l})$. Previously from the description of the response function per unit time of a scalar field (\ref{gg}), the term $\theta(\omega-m)$ suggests that momentum cannot be imaginary but now in the expression of the integrand (\ref{int}) there is no restriction of theta function on $(m^2-\omega_0^2)$, it can be either positive or negative or zero.

\begin{figure}
\includegraphics[width=8.5cm]{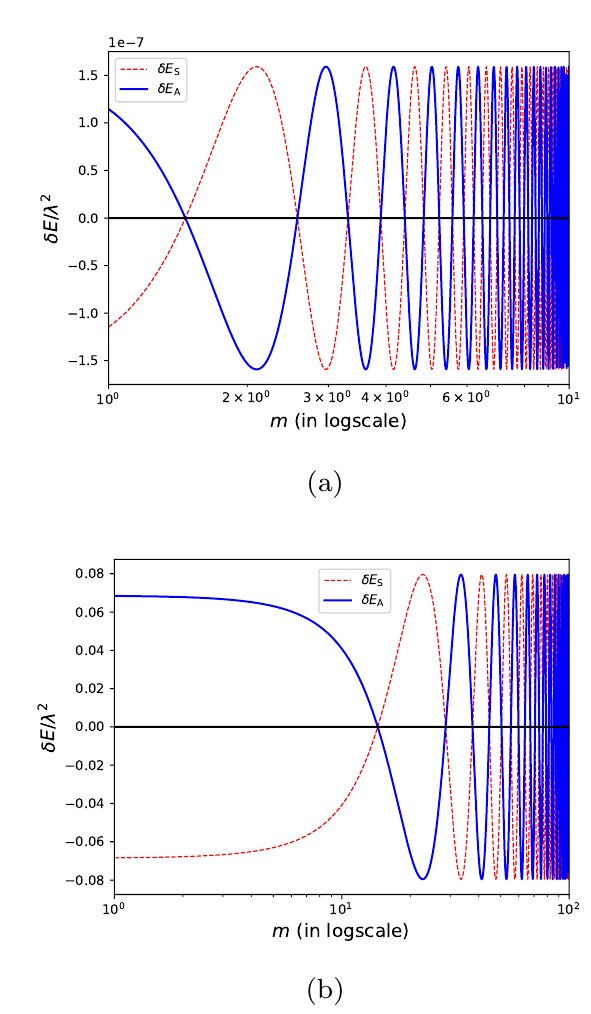}
\caption{  Dependence of energy shifts \emph{w.r.t} mass of the scalar field for $\omega_0^2>m^2$. (a) for $L>>l$  we take $L = 1000$, $l =1$, $\omega_0=10$ and $m$ is varying from 1 to 10  (b) for $l>>L$  we take $L = 1$, $l =1000$, $\omega_0=100$ and $m$ is varying from 1 to 100. Here $m,~L,~l,~\omega_0$ are in arbitrary units.
}
\label{figure1} 
\end{figure}

\begin{figure}
\vspace{.3 em}
\includegraphics[width=8.5cm]{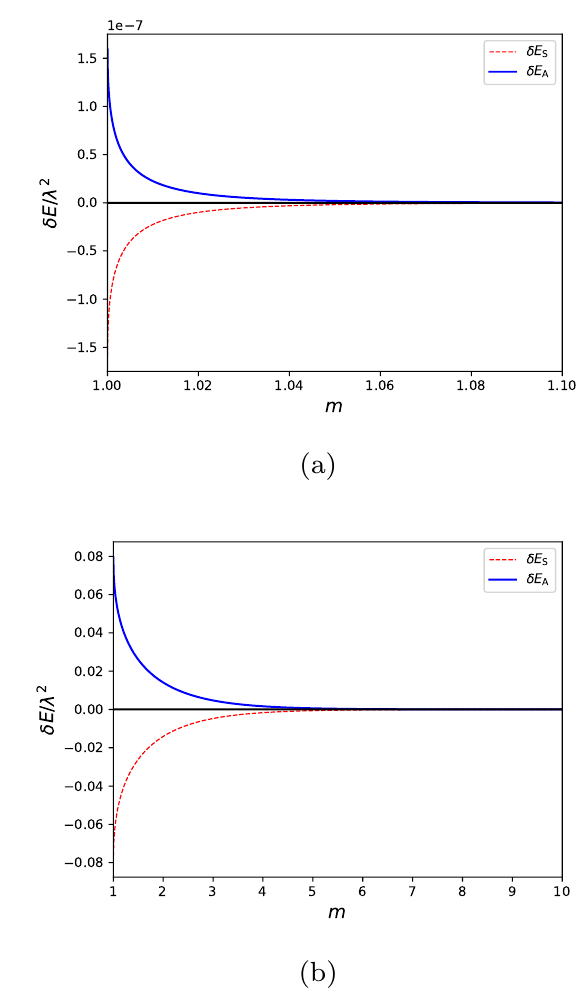}
\caption{ Dependence of energy shifts \emph{w.r.t} mass of the scalar field for $\omega_0^2<m^2$. (a) for $L>>l$  we take $L = 1000$, $l =1$, $\omega_0=1$  and $m$ is varying from 1 to 1.1 (b) for $l>>L$  we take $L = 1$, $l =1000$, $\omega_0=1$ and $m$ is varying from 1 to 10. Here $m,~L,~l,~\omega_0$ are in arbitrary units.
}
\label{figure2} 
\end{figure}

\subsection{Case-I, $\omega_0^2>m^2$}\label{case}
For case-I, the poles of the integrand (\ref{int}) lie on the real-line. Therefore,
the expressions for Lamb shift of two atoms are given by,
\begin{eqnarray}\label{interatomicinteraction}
\delta E_S &=& - \frac{\lambda^2}{4\pi L\sqrt{1+\frac{L^2}{4l^2}}}\cos(2\sqrt{\omega_0^2-m^2}\,\,l \sinh^{-1}\big(L/2l\big))~,\nn\\
\delta E_A &=& \frac{\lambda^2}{4\pi L\sqrt{1+\frac{L^2}{4l^2}}}\cos(2\sqrt{\omega_0^2-m^2}\,\,l \sinh^{-1}\big(L/2l\big)).\nn\\
\end{eqnarray}

We also note that there exists a characteristic length scale $l$, which up to the leading order 
approximation is equal to  $\frac{ \sqrt{1-\frac{r_s}{r}}}{\kappa}$, where 
$\kappa=\frac{1}{2 \rs}$ is the surface gravity \cite{singha_remarks_2019}. To analyze the behavior of the \emph{RCPI}, here we also 
consider both the limits of the proper distance between the atoms which are either
larger or smaller than the characteristic length scale.
From equation (\ref{interatomicinteraction}), it is shown that in the limit 
$L>>l$ (or $\Delta\theta>> 2 \sqrt{\frac{\delta}{r_s}}$) \cite{singha_remarks_2019}, the expressions are,
\begin{eqnarray}
\label{interatomicinteraction1} 
\delta E_{S}&=& -\frac{\lambda^2~l}{2
\pi L^2 }\cos(2 \sqrt{\omega_0^2-m^2}\,\,l
\log(L/l))~,\nonumber\\
\delta E_{A}&=&\frac{\lambda^2~l }{2
\pi L^2 }\cos(2 \sqrt{\omega_0^2-m^2}\,\,l
\log(L/l))~,
\end{eqnarray}
and in the limit $L<<l$ (or $\Delta\theta<< 2 \sqrt{\frac{\delta}{r_s}}$), 
\begin{eqnarray}
\label{interatomicinteraction2}
\delta E_{S}&=& -\frac{\lambda^2 }{4
\pi L }\cos( \sqrt{\omega_0^2-m^2}\,\, L)~,\nonumber\\
\delta E_{A}&=&\frac{\lambda^2  }{4
\pi L }\cos( \sqrt{\omega_0^2-m^2}\,\, L)~.
\end{eqnarray}
The schematic diagrams for both the limits are shown in  FIG. \ref{figure1}.

\subsection{Case-II, $\omega_0^2<m^2$} \label{casee}
For case-II, the poles in the integrand (\ref{int}) lie on the imaginary axis, the expressions for \emph{RCPI} are given by,
\begin{eqnarray}\label{interatomicinteraction3}
\delta E_S &=& - \frac{\lambda^2}{4\pi L\sqrt{1+\frac{L^2}{4l^2}}}e^{-2\sqrt{m^2-\omega_0^2}\,\, l \sinh^{-1}\big(L/2l\big)}~,\nn\\
\delta E_A &=&  \frac{\lambda^2}{4\pi L\sqrt{1+\frac{L^2}{4l^2}}}e^{-2\sqrt{m^2-\omega_0^2}\,\, l \sinh^{-1}\big(L/2l\big)}~.
\end{eqnarray}
The expressions have an exponential decaying term, which indicates the same kind of result of the two-point correlation function at $mR>>1$. Therefore, the correlation function and {RCPI} vanish in the large mass limit. 
From equation (\ref{interatomicinteraction1}), in the limit 
$L>>l$ (or $\Delta\theta>> 2 \sqrt{\frac{\delta}{r_s}}$), the expressions are given by,
\begin{eqnarray}\label{interatomicinteraction4}
\delta E_{S}&=& -\frac{\lambda^2~l}{2
\pi L^2 }e^{-2\sqrt{m^2-\omega_0^2}\,\, l 
\log(L/l)}~,\nonumber\\
\delta E_{A}&=&\frac{\lambda^2~l }{2
\pi L^2 }e^{-2\sqrt{m^2-\omega_0^2}\,\, l 
\log(L/l)}~,
\end{eqnarray}
and in the limit $L<<l$ (or $\Delta\theta<< 2 \sqrt{\frac{\delta}{r_s}}$),
\begin{eqnarray}
\label{interatomicinteraction5}
\delta E_{S}&=& -\frac{\lambda^2 }{4
\pi L }e^{- \sqrt{m^2-\omega_0^2}\,\, L}~,\nonumber\\
\delta E_{A}&=&\frac{\lambda^2  }{4
\pi L }e^{- \sqrt{m^2-\omega_0^2}\,\, L}~.
\end{eqnarray}
The schematic diagrams for both the limits are shown in FIG. \ref{figure2}~.
\subsection{Case-III, $\omega_0^2=m^2$} \label{caseee}
For case-III, the pole of the integrand (\ref{int}) is at the origin, $z=0$. The expressions for \emph{RCPI} are shown below,
\begin{eqnarray}\label{interatomicinteraction6}
\delta E_S &=& - \frac{\lambda^2}{4\pi L\sqrt{1+\frac{L^2}{4l^2}}},\nn\\
\delta E_A &=&  \frac{\lambda^2}{4\pi L\sqrt{1+\frac{L^2}{4l^2}}}.
\end{eqnarray}
From equation (\ref{interatomicinteraction3}), in the limit 
$L>>l$ (or $\Delta\theta>> 2 \sqrt{\frac{\delta}{r_s}}$), the expressions are given by,
\begin{eqnarray}\label{interatomicinteraction7}
\delta E_{S}&=& -\frac{\lambda^2~l}{2
\pi L^2 },\nonumber\\
\delta E_{A}&=&\frac{\lambda^2~l }{2
\pi L^2 },
\end{eqnarray}
and in the limit $L<<l$ (or $\Delta\theta<< 2 \sqrt{\frac{\delta}{r_s}}$),
\begin{eqnarray}
\label{interatomicinteraction8}
\delta E_{S}&=& -\frac{\lambda^2 }{4
\pi L },\nonumber\\
\delta E_{A}&=&\frac{\lambda^2  }{4
\pi L }~.
\end{eqnarray}

Both the FIG. \ref{figure1} and FIG. \ref{figure2} merge at the same point for $\omega_0=m$. We also check analytically the pair of equations, (\ref{interatomicinteraction1},\ref{interatomicinteraction4}) are equal with the expression (\ref{interatomicinteraction7}) and similarly for (\ref{interatomicinteraction2},\ref{interatomicinteraction5}) are equal with (\ref{interatomicinteraction8}). 
However, The length scale limit beyond a characteristic value ($L>>l$) for the three cases (\ref{case},\ref{casee},\ref{caseee}) is not compatible with the local flatness of the spacetime \cite{singha_remarks_2019}.

\section{Discussions}

From the analogy of \emph{equivalence principle} we can say an accelerating observer mimics an observer in a gravitational field and from \emph{thermalization  theorem} we find a connection between an accelerating atom and a quantum system connected to a thermal bath. Therefore, we use the quantum master equation approach to describe the dynamics of an accelerating two-atom system interacting with a massive scalar field. We find the expressions for the second-order shift term for system-field coupling Hamiltonian. In the presence of vacuum fluctuations of the field, the effective shift terms are responsible for the interatomic correlations between the atoms. Here we calculate the \emph{RCPI} for the Schwarzschild spacetime in the near-horizon region. From the recent studies, the \emph{RCPI} using a massless scalar field beyond a characteristic length scale follows a $1/L^2$  power-law behavior \cite{singha_remarks_2019}. The length-scale is equivalent to the inverse of the surface gravity $\kappa$ in Schwarzschild spacetime. However, the limit is not compatible with the locally flat space-time. For the massive scalar field, we also find the same kind of behavior. With this characteristic length scale, we also introduce a mass dependence of \emph{RCPI}, which serves another independent length scale in the dynamics. The key point we have addressed here is both the limits ($\omega_0^2>m^2,\omega_0^2<m^2$) merge to a constant-value which is independent of the mass of scalar field and the frequency of Zeeman levels. There is a competition between the frequency of Zeeman levels and the mass of the scalar field when they are equal the energy-shifts become maximum. This particular value exactly shows the behavior of a two-atom system whose frequency of Zeeman levels is zero in a massless scalar field. In summary, we have shown that in the low mass limit it exactly follows the behavior reported in ref. \cite{singha_remarks_2019} with the frequency modulated by a mass term and in the high mass limit, the behavior of \emph{RCPI} becomes short-range and eventually disappears beyond a characteristic length scale of $1/m$. So, the atoms become uncorrelated in the large mass limit. 

\begin{acknowledgments}

We sincerely thank Rangeet Bhattacharyya, Golam Mortuza Hossain, Sumanta Chakraborty, Siddharth Lal, Bibhas Ranjan Majhi, Ashmita Das, Arnab Chakrabarti and Chandramouli Chowdhury for many 
useful discussions. AC and SS thank University Grant Commission (UGC) for a junior research fellowship.

\end{acknowledgments}

\end{document}